\documentclass[epsf,aps,preprintnumbers,amsmath,amssymb,11pt]{revtex4}

\usepackage{epsf}

\usepackage{graphicx}

\usepackage{bm}

\begin {document}

%%%%%%%%%%%%%%%%%%%%%%%%%%%%%%%%%%%%%%%%%%%%%%%%%%%%%%%%%%%%%%%%%%%%%%%%%%%%%%%

%\def\Re{\operatorname{Re}}

\def\be{\begin{equation}}
\def\ee{\end{equation}}
\def\bea{\begin{eqnarray*}}
\def\eea{\end{eqnarray*}}

%%%%%%%%%%%%%%%%%%%%%%%%%%%%%%%%%%%%%%%%%%%%%%%%%%%%%%%%%%%%%%%%%%%%%%%%%%%%%%%

\title
    {
Tachyonic crystals and the laminar instability of the perturbative vacuum \\in asymptotically free gauge theory
    }

\author {H. B. Thacker}
\affiliation
    {%
 Department of Physics,
    University of Virginia,
    P.O. Box 400714
    Charlottesville, VA 22901-4714\\
}

\date{\today}

\begin {abstract}%
Lattice Monte Carlo studies in $SU(3)$ gauge theory have shown that the topological charge distribution in 
the vacuum is dominated by thin coherent membranes of codimension one arranged in a layered, alternating-sign sandwich. 
A similar lamination of topological charge occurs in the 2D $CP^{N-1}$ model. In holographic QCD, the observed topological
charge sheets are naturally interpreted as $D6$ branes wrapped around an $S_4$.. With this interpretation, the laminated array of topological charge membranes
observed on the lattice can be identified as a ``tachyonic crystal,'' a regular, alternating-sign 
array of $D6$ and $\bar{D6}$ branes that arises as the final state of the
decay of a non-BPS $D7$ brane via the tachyonic mode of the attached string. In the gauge theory, the homogeneous, 
space-filling $D7$ brane represents the perturbative gauge vacuum, which is unstable toward lamination associated with
a marginal tachyonic boundary perturbation $\propto \cos(X/\sqrt{2\alpha'})$. For the $CP^{N-1}$ model, the cutoff field theory can be cast as the low energy limit of an open string
theory in background gauge and tachyon fields $A_{\mu}(x)$ and $\lambda(x)$. This allows a detailed comparison with large $N$ field theory results and provides
strong support for the tachyonic crystal interpretation of the gauge theory vacuum.
    {%

. }% 
\end {abstract}

\maketitle
\thispagestyle {empty}

 %%%%%%%%%%%%%%%%%%%%%%%%%%%%%%%%%%%%%%%%%%%%%%%%%%%%%%%%%%%%%%%%%%%%%%%%%%%%%%%

\section {Introduction}
\label{sec:intro}

Monte Carlo studies of 4-dimensional SU(3) gauge theory \cite{Horvath03,Horvath_global} and parallel studies of 2-dimensional $CP^{N-1}$ models \cite{Ahmad05,Lian07,Keith-Hynes08} have shown that the
topological charge density in the vacuum of these theories is dominated by extended coherent codimension-one sheets.
These results point to a new paradigm for the physical vacuum of asymptotically
free gauge theories. The vacuum appears to be a laminated ``topological sandwich'' consisting of densely packed, alternating sign membranes of topological charge. 
The possible interpretation of these topological
charge membranes as the holographic dual of D6 branes in Witten's type IIa string theory formulation of holographic QCD \cite{Witten98} has been discussed in \cite{Ahmad05,lat06}.  Locally, the membranes 
are flat, parallel, alternating in sign, and roughly equally spaced. Over larger distances they bend and fold such that the local 
orientation of the topological sandwich decorrelates over distances large compared to the confinement scale. 
(This last statement is based mainly on visual inspection of $CP^{N-1}$ gauge configurations for various correlation lengths.)

\begin{figure}
   \begin{center}
     \vskip -0.15in
     \centerline{
     \includegraphics[width=9.5truecm,angle=-90]{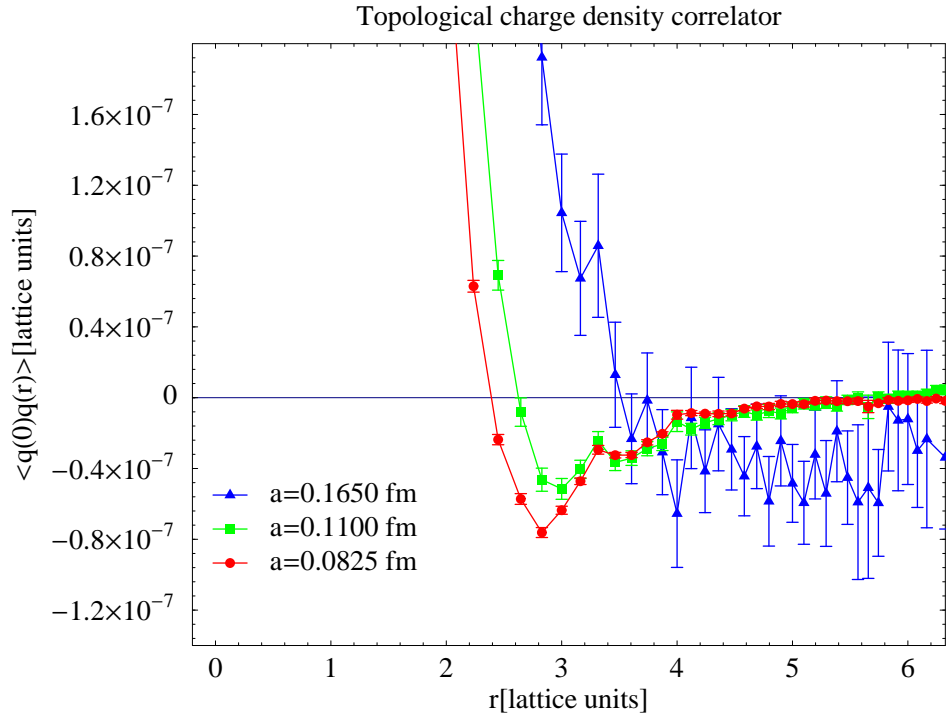}
     }
     \vskip 0.15in
     \caption{The two-point topological charge correlator in pure glue QCD. The horizontal axis is in lattice units and results are shown
for 3 values of physical lattice spacing $a$.}
     \label{fig:qcd_2point}
   \end{center}
\end{figure}

\begin{figure}
   \begin{center}
     \vskip -0.15in
     \centerline{
     \includegraphics[width=9.5truecm,angle=-90]{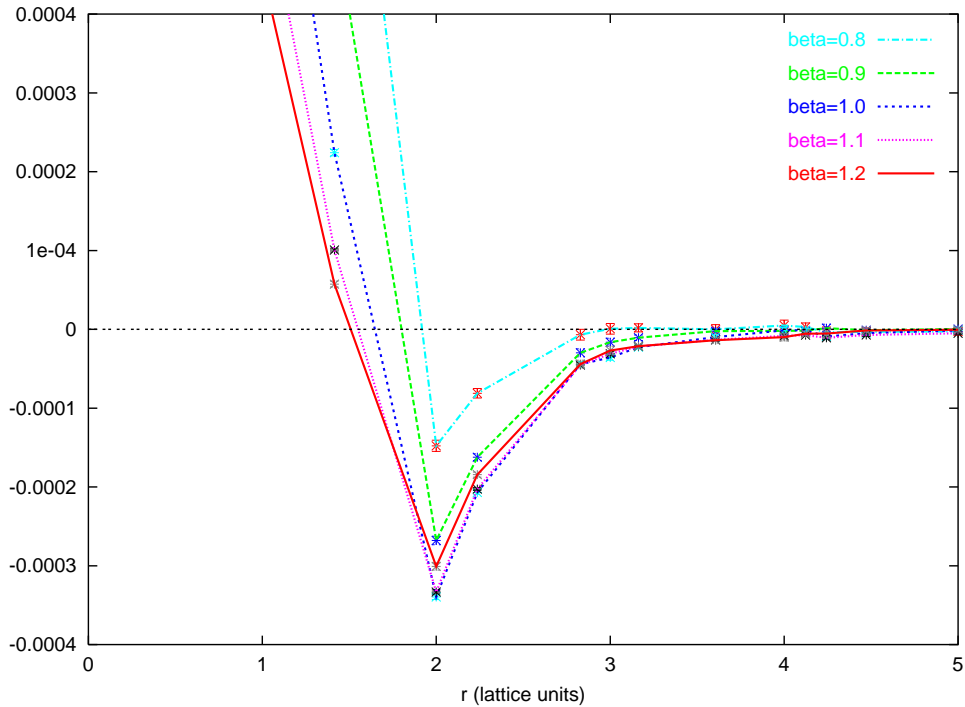}
     }
     \vskip 0.15in
     \caption{The two-point topological charge correlator for the 2-dimensional $CP^3$ sigma model. The correlation length determined from the $\bar{Z}Z$ meson mass
varies by a factor of 5 over the range of $\beta$'s plotted.}
     \label{fig:cpn_2point}
   \end{center}
\end{figure}

An important characteristic of the observed  charge membranes is that both the transverse thickness of the branes and the typical 
separation between adjacent, oppositely charged branes is roughly constant in lattice units, of order a few lattice spacings. 
This basic structure leads to the characteristic behavior of the topological charge two-point
correlator, which consists of a narrow positive peak at $x\approx 0$ and a negative tail with a pronounced
negative dip at a separation of around 2 lattice spacings as shown in Fig. \ref{fig:qcd_2point} for 4-dimensional $SU(3)$ 
gauge theory and in Fig. \ref{fig:cpn_2point} for the 2-dimensional $CP^3$ model. 
The position of the negative dip remains roughly the same in lattice units over a wide range of correlation
lengths. Thus, both the thickness and the typical spacing of the topological charge sheets goes to zero with the lattice spacing in the 
continuum limit. At first one might wonder if structure that varies rapidly
at the cutoff scale can or should influence long range physics in the continuum limit. But it is easily seen numerically that the membranes are responsible for building up an {\it expected} positive
delta-function contact term in the topological charge correlator \cite{Seiler87}. The contribution of this contact term to the topological susceptibility $\chi_t$, (given by the integrated correlator)
does not go away in the continuum limit, but in fact combines with the contribution of the negative tail at $|x|>0$ to give a positive topological susceptibility which scales properly
with the mass gap. (This is true for 4-dimensional $SU(3)$ gauge theory and 2-dimensional $CP^{N-1}$ for $N>3$. 
The only exceptions are the $CP^1$ and $CP^2$ models, which are dominated by small instantons and do not exhibit proper scaling for $\chi_t$ \cite{Berg,Luscher82,Lian07}.) 
The laminated structure appearing at the cutoff scale is thus essential for obtaining the proper continuum physics. Among other things, the topological sandwich vacuum serves as a 
medium which supports the physical fluctuations of topological charge density in the vacuum, which arise from the bends, folds, phonons, and other deviations from a regular one-dimensional crystal structure.
This has the notable effect of replacing locally quantized instanton fluctuations, which would exist in the perturbative vacuum, with unquantized gaussian fluctuations. (Topological charge can
flow to and from infinity in a continuous way along the membranes.) The relaxation of the local quantization constraint 
causes the $\theta$-dependence of the vacuum energy to change from the instanton gas prediction $\propto 1-\cos\theta$
to the large N prediction $\propto \theta^2$ \cite{Keith-Hynes08}.

The Monte Carlo evidence suggests that, with a finite cutoff, the perturbative vacuum of an asymptotically free confining gauge theory has an instability toward lamination along
some direction, and that this lamination characterizes the transition from the perturbative vacuum to the physical, confining vacuum. In this paper I will explore the remarkable similarity between the
layered, alternating-sign topological sandwich structure observed in the Monte Carlo studies and a well-known but somewhat mysterious structure in string theory 
called a ``tachyonic crystal'' \cite{Polchinski94}. This is a layered, alternating sign, one-dimensional
array of codimension-one $Dp$ branes, which arises from the tachyonic decay of an unstable $D(p+1)$ brane.
In the IIa formulation of holographic QCD, the topological sandwich of D6 branes can be intepreted as the stable tachyonic crystal which arises from the decay of
a single spacetime filling $D7$ brane. (Both the $D6$ branes and the $D7$ brane are wrapped around a compact $S_4$ and thus appear as 3- and 4-dimensional objects, respectively, in
4-dimensional spacetime.) In this framework, the homogeneous but unstable $D7$ brane embodies the axial $U(1)$ symmetry associated with the $\theta$-parameter of the gauge theory.  
Its decay to an array of $D6$ branes breaks this symmetry, thus implementing the axial $U(1)$ anomaly and producing a physical vacuum with nonzero topological susceptibility. 
The boundary conformal field theory (BCFT) methods
which provide the framework for our understanding of unstable brane decay \cite{Sen02} should thus be directly applicable to the description of the physical vacuum in asymptotically free
gauge theories.

In order to gain a clearer understanding of the formation and properties of the topological sandwich vacuum, we will explore these issues in the 
2-dimensional $CP^{N-1}$ model, where we already have some understanding of the physical vacuum based on large N analysis. Since they were first introduced, the 
$CP^{N-1}$  models have served as an instructive example of topological charge structure which is in many
ways quite similar to 4-dimensional $SU(N)$ gauge theory. The study of topological charge distributions in Monte Carlo generated $CP^{N-1}$ gauge configurations
has revealed a topological sandwich structure of extended coherent codimension-one objects \cite{Ahmad05}, completely analogous to the structure found in QCD. 
The $CP^{N-1}$ model may be formulated as a theory of N complex scalar fields $Z_i(x),\; i=1,\ldots, N$, interacting with an auxiliary $U(1)$ gauge field $A_{\mu}(x)$ and a scalar Lagrange multiplier
field $\lambda(x)$ (which incorporates the constraint $Z^{\dag} Z=1$). In the large N approximation (with a momentum space cutoff), the $\lambda$ field acquires a VEV, which
gives a mass to the $Z$ particles. With massive $Z$'s, the one loop effective action for the gauge field acquires an $(F_{\mu\nu})^2$ term, producing a linear Coulomb potential
which confines $U(1)$ charge. 
We will show, by writing the Z-particle propagator in a proper time representation, that the $CP^{N-1}$
model can be obtained from the $\alpha'\rightarrow 0$ limit of a bosonic open string theory with one end interacting with a background gauge field $A_{\mu}(X)$ and tachyon field
$\lambda(X)$. In this formulation of the model, the string length $\sqrt{\alpha'}$ serves as a short distance cutoff and plays a role analogous to the lattice spacing in the
lattice formulation of the cutoff theory. With this string theoretic cutoff, we can relate the dynamics of the $CP^{N-1}$ model to that of a
boundary conformal field theory (BCFT), where the tachyon background $\lambda(X)$ admits an exactly marginal, oscillatory boundary perturbation of the form \cite{Callan94,Polchinski94}
\begin{equation}
\label{eq:marginal}
\lambda(X) = -\frac{1}{2}\left[g\exp\left(\frac{iX(0)}{\sqrt{2\alpha'}}\right) + \bar{g}\exp\left(-\frac{iX(0)}{\sqrt{2\alpha'}}\right)\right]
\end{equation}
Here $X(0)$ is the endpoint coordinate of an open bosonic string, given by the boundary value on the real axis of holomorphic and antiholomorphic conformal fields in
the upper half plane. 
The phase of the parameter $g$ fixes the overall position of the oscillating field along the $X$ axis.  From here on we will take $g$ to be real and positive.
Sen has shown that this  boundary perturbation (\ref{eq:marginal}) represents the Euclidean world sheet view of the homogeneous decay of an unstable D-brane \cite{Sen02}.
The theory without a boundary term ($g=0$) corresponds to Neumann boundary conditions on the $X$ coordinate, i.e. a homogeneous D-brane spanning that coordinate.
For $0<g<\frac{1}{2}$, the perturbation (\ref{eq:marginal}) is marginal and belongs to an SU(2) current algebra. As we discuss in the following Section, this boundary deformation preserves world sheet
conformal symmetry, and the open string spectrum is obtained in terms of $U(1)$ Kac-Moody modules of a ``phase shifted'' Sugawara Hamiltonian. 
At a critical value $g=\frac{1}{2}$ the b.c.'s on $X$ become pure Dirichlet conditions. In space-time, this represents the background state reached after the  brane decay process is complete,
consisting of an alternating-sign codimension-one array of D-branes arranged along the $X$ direction in a one-dimensional tachyonic crystal \cite{Gaiotto03,Lambert07}.

\section{String theory as an ultraviolet completion of the $CP^{N-1}$ model}

The $CP^{N-1}$ model is a nonlinear sigma model of $N$ self-interacting complex scalar fields $Z_i(x),\; i=1,\ldots,N$
with the constraint $Z^{\dag} Z=1$. The most convenient formulation of the model for our purposes is obtained 
by introducing an auxiliary abelian gauge field $A_{\mu}(x)$. Then the action consists solely of a gauge 
covariant kinetic term,
\begin{equation}
\label{eq:lagrangian}
{\cal L} = \frac{N}{\gamma^2}\int d^2x\left(D^{\mu}Z\right)^{\dag}\left(D_{\mu}Z\right)
\end{equation}
Here we will denote the $CP^{N-1}$ coupling constant by $\gamma$ (rather than the more conventional $g$) to avoid confusion with the coefficient of the boundary perturbation.
In (\ref{eq:lagrangian}) the covariant derivative is $D_{\mu}=\partial_{\mu}-iA_{\mu}$. Note that there is no $(F_{\mu\nu})^2$ term for the gauge field, so $A_{\mu}$ is a dummy field 
whose equation of motion sets it equal to the $U(1)$ current,
\begin{equation}
\label{eq:U1current}
A_{\mu} = Z^{\dag}(\partial_{\mu}Z) - (\partial_{\mu}Z)^{\dag}Z
\end{equation}
If the $A_{\mu}$ field is integrated out, the theory can be written as a nonlinear sigma model of self-interacting $Z$'s.
Because of the absence of a kinetic term for the gauge field, the model is classically conformally invariant. It acquires a mass scale via dimensional transmutation, as in 4-dimensional QCD.
An $(F_{\mu\nu})^2$ term for the gauge field is generated by loop effects, as can be seen explicitly in the large $N$ solution \cite{Dadda78,Witten79}. This produces a linear Coulomb potential which
confines $U(1)$ charge. The generation of a kinetic term in the effective gauge action is a direct consequence of the dynamical generation of a mass for the $Z$-particles. In large N, the 
transition from the perturbative to the physical vacuum is constrained by a one-loop consistency condition which determines the ``constituent'' mass of the $Z$ particles.
In our string-cutoff theory, we will show that this transition is described in terms of the marginal boundary perturbation (\ref{eq:marginal}) in the  BCFT for a massless scalar field
on the string world sheet. 

To clarify the connection between the $CP^{N-1}$ model and boundary CFT, it is useful to introduce a proper time formalism. First let us impose the constraint
$Z^{\dag} Z=1$ via an auxilary Lagrange multiplier field,
\begin{equation}
{\cal L}\rightarrow {\cal L} -\lambda(x)(Z^{\dag}Z-1)
\end{equation}
Thus the model reduces to charged scalar particles interacting with a real scalar field $\lambda(x)$ and a vector field $A_{\mu}(x)$. We will focus on the two-point function for
$Z$ particles,
\begin{equation}
\langle Z_i(x_1)Z_j^{\dag}(x_2)\rangle = \frac{1}{N}\delta_{i j}G(x_1,x_2)
\end{equation}
By introducing a proper time variable, we can replace the path integral over the $Z$ fields by an integral over spacetime particle paths $x^{\mu}(\tau)$, $\mu=1,2$. (We work in 
two-dimensional Euclidean space.) The exact propagator for a charged particle with fixed $\lambda$ and $A_{\mu}$ fields is then given in a proper time representation,
\begin{equation}
G(x_1,x_2) = \int_0^{\infty} \hat{G}(x_1,x_2,T) \;dT
\end{equation}
where the Green's function for the proper time Schrodinger equation is given by summing over particle paths with boundary conditions $x^{\mu}(0)=x_1,\; x^{\mu}(T)=x_2$.
The interaction with the gauge field is given by the Wilson line along each path,
\begin{equation}
\label{eq:propertime}
\hat{G}(x_1,x_2,T) = \int {\cal D}x\; \exp i\left[\int_0^T d\tau \left(\left(\frac{dx^{\mu}}{d\tau}\right)^2 - \lambda(x)\right)-\int d\tau\frac{dx^{\mu}}{d\tau}A_{\mu}(x)\right]
\end{equation}
[Note that to get the correct equations of motion, the action should also include a metric factor $e^{-1}(\tau)$ to incorporate invariance under reparametrizations of the path $x(\tau)$.
For our discussion, we have fixed the gauge by setting $e(\tau)=1$.]

By performing the path integral over the $\lambda$ and $A_{\mu}$ fields (with the effective gauge action generated by closed $Z$-loops), we obtain the full two-point function for the model.
So far we have not explicitly introduced an ultraviolet cutoff. In a continuum formulation of the large $N$ approximation
an ordinary momentum space cutoff can be used. A lattice formulation of the $CP^{N-1}$ model provides a more general nonperturbative cutoff procedure,
but, as discussed in the Introduction, Monte Carlo results raise subtle issues regarding the scaling behavior of the topological susceptibility $\chi_t$ in the continuum limit of the lattice 
cutoff \cite{Berg,Luscher82,Ahmad05,Lian07}. The dominance of topological charge structures with transverse size of order the lattice spacing (pointlike instantons for small $N$ 
and line-like excitations for $N>4$ \cite{Lian07})
clearly indicates that the structure of the nonperturbative vacuum is crucially affected by gauge field dynamics taking place at or near the cutoff scale. 
In this paper we explore this dynamics by introducing a cutoff at short proper time, $\tau\approx 0$ in Eq. (\ref{eq:propertime}) \cite{Schwinger51}. As a nonperturbative regulator, the proper time
cutoff procedure provides a viable alternative to the lattice formulation of the model. Moreover, it leads rather naturally to a string-theoretic ultraviolet completion
in which the $CP^{N-1}$ field theory is obtained as the $\alpha'\rightarrow 0$  limit of an open string theory.
In the string theory completion of this model, the parameter $\sqrt{\alpha'}$ plays a role analogous to lattice spacing in the lattice cutoff. In the field theory limit the strings become very short,
and the world sheet parametrization of the string coordinates $X^{\mu}(\tau,\sigma)$ reduces to the proper time parametrization of the particle path $x^{\mu}(\tau)$.
The open string propagator reduces to the propagator for the $CP^{N-1}$ field $Z(x)$. For the full string theory, the contribution of massive string excitations to the propagator at a given proper time
is determined by an elliptic modulus parameter $q=e^{-\tau/4\alpha'}$. For $\tau>>4\alpha'$, only the lowest mass states contribute significantly (including the tachyon, which is associated with 
the vacuum instability as we discuss below).
Thus, in the proper time decomposition of the propagator, the stringy corrections affect the theory only at proper time separations comparable to or smaller than $\alpha'$, 
and we can regard the ``string length'' $l=\sqrt{\alpha'}$ as
a short distance cutoff. It is important for the following analysis that both the Lagrange multiplier field $\lambda$ and the gauge field $A_{\mu}$ are introduced 
as {\it boundary terms} (rather than bulk terms) on the open string world sheet. In this way, the spacetime conformal symmetry is broken but conformal symmetry on the
world sheet may be preserved by certain boundary fields. As is familiar in string theory, the boundary perturbations that preserve worldsheet conformal symmetry
correspond to background fields which satisfy the spacetime equations of motion. We will use this connection in Section V to obtain the effective space-time gauge action
and compare it with the large $N$ result.

From the expression (\ref{eq:propertime}) for the two-point function we determine the path- and field-dependent action for $Z$-particle propagation in the $CP^{N-1}$ model
to be (up to an overall constant which determines the scale of the proper time parameter $\tau$)
\begin{equation}
\label{eq:pathaction}
S \propto \int_0^T d\tau \left(\left(\frac{dx^{\mu}}{d\tau}\right)^2 - \lambda(x)-\frac{dx^{\mu}}{d\tau}A_{\mu}(x)\right)
\end{equation}
The string theoretic cutoff procedure described above consists of identifying the path action (\ref{eq:pathaction}) as the $l\rightarrow 0$ limit of the worldsheet action
for an open bosonic string coordinate $X(\sigma,\tau)$,
\begin{equation}
\label{eq:worldsheetaction}
S_{string} = S_{bulk} + S_{boundary}
\end{equation}
The bulk term is the usual CFT for a bosonic string coordinate, defined inside a rectangular area $R$ of width $l$ and length $T$ in the complex $z\equiv \tau+i\sigma$ plane.
\begin{equation}
S_{bulk} = \int_0^T d\tau \int_0^l d\sigma\left(\left(\frac{dX^{\mu}}{d\tau}\right)^2 + \left(\frac{dX^{\mu}}{d\sigma}\right)^2\right) 
= \int_R \;d^2z \partial_a X^{\mu} \partial^a X_{\mu}
\end{equation}
The worldsheet action also includes two boundary terms, corresponding to the spacetime fields $\lambda(X)$ and $A_{\mu}(X)$ interacting with one or both ends of the string,
The most physically sensible formulation for our purposes is to put the $Z$-particle charge that interacts with the background fields at one end of the string ($\sigma=0$) and impose Neumann boundary 
conditions at the other end ($\sigma=l$). (Physically, the charged particle still follows a definite path $x_{\mu}(\tau)=X_{\mu}(0,\tau)$ but now has a string attached to it.)
Thus we take the boundary term to be along the real axis,
\begin{equation}
\label{eq:boundary}
S_{boundary} = \int_0^T d\tau \left[
\lambda(X)+\frac{dX^{\mu}}{d\tau}A_{\mu}(X)\right]_{\sigma=0}
\end{equation}
(\ref{eq:worldsheetaction})-(\ref{eq:boundary}) defines the worldsheet action for the coordinates of an open bosonic string in background tachyon and U(1) gauge fields. 

The string theoretic cutoff for the $CP^{N-1}$ model may also be understood by writing the one-loop vacuum amplitude for an open string in a spectral representation
which sums over intermediate open  string states. This spectral analysis also exposes the essential role of the open string tachyon, and its connection to the instability of the $\lambda=0$
vacuum seen in the large $N$ field theory. Let us begin by considering the one-loop vacuum string amplitude $G(x,x)$ without background fields (i.e. $\lambda=A_{\mu}=0$). This amplitude has the form
\begin{equation}
\label{eq:stringprop}
G = {\rm const.}\times \int\frac{d^2p}{(2\pi)^2}\sum_i\frac{1}{p^2+m_i^2} = {\rm const.}\times \int\frac{d^2p}{(2\pi)^2}\int d\tau\;e^{-\tau p^2} Z(e^{-\tau/4\alpha'})
\end{equation}
where
\begin{equation}
\label{eq:partition}
Z(e^{-\tau/4\alpha'}) \equiv \sum_{i=0}^{\infty}e^{-\tau m_i^2}
\end{equation}
Here the sum is over all open string states satisfying the specified boundary conditions, and we consider only the motion of the string coordinates in 2-dimensional Euclidean spacetime.
With Neumann boundary conditions and in the absence of background fields, it is straightforward to separate the Virasoro eigenvalues into a $p^2$ term from translational motion
and an $m_i^2$ term from internal motion. For general background fields, such a separation is not possible, but we will consider a background $\lambda(X)$ field that depends 
on only one spatial coordinate $X$, and is periodic in that coordinate, $\lambda(X)=\lambda(X+2\pi R_0)$, where $R_0 = \sqrt{2\alpha'}$ is the self-dual radius. For
this case, the open string spectrum exhibits a band structure that can be understood in terms of Bloch wave states, as discussed in the next Section.
[In the case of HQCD, the justification for considering solutions which depend on only one spacetime coordinate follows from the observation that, while the $D7$ brane, 
which represents the perturbative vacuum, is non-BPS and unstable to tachyonic decay, the D6 branes which result from 
lamination along one coordinate direction are BPS and stable against further dimensional collapse.]

For the free string with $\lambda=0$, a crucial feature of the sum over discrete string states (\ref{eq:partition}) is that it yields an amplitude
with modular properties characterized by an elliptic nome 
\begin{equation}
w=e^{-\tau/4\alpha'}
\end{equation}
For a single compactified bosonic coordinate at the radius $R=R_0$ satisfying Neumann boundary conditions at both ends of the string, the sum over states in (\ref{eq:partition}) gives
\begin{equation}
\label{eq:Z_NN}
Z^{R_0}_{NN}(w) \equiv \sum_i e^{-\tau m_i^2} = \frac{w^{-\frac{1}{24}}}{f(w)}\sum_{n=-\infty}^{\infty} w^{n^2}
\end{equation}
where $f(w) = \prod_{n=1}^{\infty}(1-w^n)$. The factor $w^{-\frac{1}{24}}$ reflects the negative mass squared of the tachyon mode. Physically, this mode arises as a Casimir effect
from the zero-point energy of the open string oscillators. The presence of a tachyon mode in the string-theoretic completion of the
$CP^{N-1}$ model is a signal that the perturbative vacuum in the cutoff field theory is unstable toward tachyon condensation. Note that the mass of the tachyon is of order the 
cutoff scale, just like the positive mass-squared states of the string. However, unlike the positive states, the tachyon does not decouple in the continuum limit, but in fact 
drives the transition from the perturbative vacuum to the physical vacuum. The scenario for this transition follows Sen's description of the tachyon-induced decay of an unstable D-brane \cite{Sen02}. 
Boundary conformal field theory for the type IIA string in D-brane backgrounds should play a role in the description of the QCD vacuum analogous to the treatment of 
the $CP^{N-1}$ model that we consider in this paper. In both cases, we are dealing with the laminar collapse of a spacetime filling D-brane which occurs locally along
a single coordinate direction, with the BCFT for that coordinate acquiring a tachyonic boundary perturbation. 

\section{Bloch waves and band structure for a propagating string: the connection between $R=R_0$ and $R=\infty$}

For the oscillatory boundary perturbation 
\begin{equation}
\label{eq:marginal2}
\lambda(X) = g \cos (X/\sqrt{2\alpha'})
\end{equation}
the spectrum and partition function for the bosonic string have been analyzed both for the case $R=R_0$ and $R=\infty$ in Refs. \cite{Callan94,Polchinski94}.
Here we are interested in the bulk properties of the zero temperature $CP^{N-1}$ model, so the case $R=\infty$ is most directly relevant to our discussion. However,
it is instructive to first review the properties of the string spectrum with coordinate compactified at the $SU(2)$ radius $R_0$. The spectrum at $R=\infty$
can then be understood in terms of Bloch wave states. The $R=\infty$ theory exhibits a ``stringy'' band structure, with each band corresponding to one of the eigenstates of
the compactified $R=R_0$ theory \cite{Polchinski94,Callan94}. For the $R=\infty$ theory with boundary perturbation (\ref{eq:marginal2}), 
the Bloch wave basis allows a separation between translational and internal motion analogous to the free string case.
Note that the spacing between branes is set 
by the periodicity of the boundary perturbation (\ref{eq:marginal2}), $2\pi R_0$. This period is also the circumference of the compactified coordinate at the $SU(2)$ radius, so 
the theory at the $SU(2)$ compactification point describes a unit cell of the tachyonic crystal in the $R=\infty$ theory.  

For nonzero $g$, the eigenvalue spectrum of the $R=\infty$ theory consists of 
bands separated by gaps. Just as in the case of an ordinary particle in a periodic crystal, there are two complementary viewpoints to describe the spectrum of bands and band gaps, the tight binding
approximation (TBA) and the nearly-free string (NFS) approximation. In the TBA, the Bloch wave states of the $R=\infty$ theory are constructed from plane wave superpositions of eigenstates
defined on a unit cell. Each eigenstate on the unit cell gives rise to a band of eigenvalues for the $R=\infty$ theory. The critical boundary perturbation $g=\frac{1}{2}$
corresponds to the tight binding limit, in which the states in a band become degenerate, and the spectrum for $R=\infty$ is just a replication, on each cell, of the discrete spectrum at the SU(2) radius.
This is the case when the end of the string is firmly stuck to a single D-brane of the tachyonic crystal and does not hop to neighboring cells, so it behaves the same as in the
compactified theory. 
For subcritical boundary perturbations, $g<\frac{1}{2}$, propagating states and momentum dependence for the energy eigenvalues within a band arise from mixing between the states on neighboring unit cells. On the other hand, for $g \approx 0$,
the NFS view of the band structure starts with the continuous relativistic dispersion relation of the free string at $R=\infty$. The boundary perturbation injects discrete units of momentum which are integer
multiples of $R_0^{-1}$, causing left- and right-propagating string states
near the edges of the Brillouin zones to mix, opening up band gaps at the zone boundaries. 

To obtain the spectrum of the compactified $R=R_0$ theory, define the $SU(2)$ Kac-Moody currents constructed from the holomorphic (left-moving) part of the coordinate $X_L(z)$ (where $z=\tau+i\sigma$),
\begin{equation}
J_3= \frac{i}{\sqrt{2}}\partial X_L, \;\; J^{\pm} \equiv J_1 \pm iJ_2 = e^{\pm i\sqrt{2}X_L}
\end{equation}
There is also a set of right-moving Kac-Moody currents, but by standard arguments, the open string Hamiltonian can be written entirely in terms of the left-moving fields defined on
the doubled line segment $-l<\sigma<l$ with periodic boundary conditions. The free open string Hamiltonian density is given by a Sugawara construction which, by $SU(2)$ symmetry, can be written
in several ways
\begin{equation}
{\cal H}_0(\sigma) = :J^1 J^1: = :J^3J^3: = \frac{1}{3}:{\bf J}\cdot{\bf J}:
\end{equation}
In terms of the Kac-Moody currents, the marginal boundary perturbation (\ref{eq:marginal2}) is simply
\begin{equation}
{\cal H}_{int}=\lambda(X(\sigma=0)) = g J^1(\sigma=0)
\end{equation}
For the free string in the $:J^3 J^3:$ form, the partition function is easily calculated by observing that the states fall into $U(1)$ Kac-Moody modules, one for each allowed value of the $U(1)$ charge 
$Q= \int d\sigma/(2\pi) J^3(\sigma)$. The module with conformal weight $Q^2$ contributes $w^{-1/24}f(w)^{-1}w^{Q^2}$ to the partition function. The Kac-Moody charges are equal to the allowed
translational zero mode momenta of the string, which for the $R=R_0$ compactified theory are integers. The sum over modules gives the free string partition function,
\begin{equation}
\label{eq:Z_R0}
Z_{NN}^{R_0} = \frac{w^{-\frac{1}{24}}}{f(w)}\sum_{Q\epsilon {\bf Z}} w^{Q^2}
\end{equation}
  
By $SU(2)$ symmetry, the same partition function is obtained from the $:J^1 J^1:$ form of the Sugawara Hamiltonian. In this latter form, the marginal boundary perturbation
can be included simply by shifting all of the Fourier modes of $J^1$ by a constant,
\begin{equation}
{\cal J}^1_n \equiv J^1_n +\frac{1}{2}g
\end{equation}
The Hamiltonian including the boundary interaction is still of Sugawara form,
\begin{equation}
{\cal H}_0 + {\cal H}_{int} = :{\cal J}^1{\cal J}^1: + {\rm const.}\;\;,
\end{equation}
and the partition function is still given by a sum over $U(1)$ Kac-Moody modules, but with weights shifted by a constant due to the boundary interaction,
\begin{equation}
Z_{BN}^{R_0}  = \frac{w^{-\frac{1}{24}}}{f(w)}\sum_{Q=-\infty}^{\infty} w^{(Q+\frac{g}{2})^2}
\end{equation}

The band structure of the $R=\infty$ spectrum was discussed in Refs. \cite{Polchinski94,Callan94} for 
the case in which both ends of the string interact with the tachyonic background with equal strength. 
Here we will derive the corresponding results for the case of interest
in our string-cutoff $CP^{N-1}$ model, where only one end of the string interacts with the tachyon background. 
By open string/closed string duality, the partion function for the open string can
be rewritten as a partition function for a closed string propagating from one boundary to the other, 
\begin{equation}
 Z_{BN}^{R=\infty} = \langle B|q^{L_0+\bar{L}_0}|N\rangle_{R=\infty}
\end{equation}
Here $|N\rangle$ is the CFT boundary state associated with Neumann boundary conditions at the string end, and $|B\rangle$ is the same state ``rotated''
by the boundary perturbation. $q$ is the conjugate nome related to $w$ by a Jacobi transformation,
\begin{equation}
\log q = \frac{2\pi^2}{\log w}
\end{equation}
By writing the boundary states in terms of closed string left- and right-moving $SU(2)$ Kac-Moody modules and
observing that the boundary perturbation (\ref{eq:marginal2}) is a global $SU(2)$ rotation acting on the left-moving states, we find
\begin{eqnarray}
\label{eq:Z_BN1}
Z_{BN}^{R=\infty} = \frac{1}{\sqrt{2}}\sum_{j=0,\frac{1}{2},1,\ldots} \chi_j^{Vir}(q^2)\int_{-\pi}^{\pi} \frac{dp}{2\pi}\sum_{m=-j}^j\langle j,m|e^{ipJ_3}e^{i\pi gJ_1}e^{ipJ_3}|j,m\rangle \nonumber \\
=\frac{1}{\sqrt{2}}\sum_{j=0,\frac{1}{2},1,\ldots} \chi_j^{Vir}(q^2)\int_{-\pi}^{\pi} \frac{dp}{2\pi}\sum_{m=-j}^j\frac{\sin(2j+1)\beta/2}{\sin \beta/2}
\end{eqnarray}
Here $\chi_j^{Vir}(q)= q^{-\frac{1}{24}}f(q)^{-1}(q^{j^2}-q^{(j+1)^2})$ is the discrete Virasoro character of spin $j$, and the $J_i$'s are the $SU(2)$ generators in the closed string channel.
The net rotation angle $\beta$ is given in terms of the momentum $p$ and the boundary perturbation $g$ by
\begin{equation}
\label{eq:angle}
\cos(\beta/2) = \cos (\pi g) \cos p
\end{equation}
Finally, we can write the partition function in terms of open string variables using the OS/CS duality identity:
\begin{equation}
\label{eq:OSCS}
\frac{1}{\sqrt{2}}\sum_{j=0,\frac{1}{2},1,\ldots}\chi_j^{Vir}(q^2)\frac{\sin(2j+1)\beta/2}{\sin \beta/2}=\frac{w^{-\frac{1}{24}}}{f(w)}\sum_{n=-\infty}^{\infty}w^{(n+\beta/4\pi)^2}
\end{equation}
This gives the partition function
\begin{equation}
\label{eq:R_infty}
Z_{BN}^{R=\infty} = \int_{-\pi}^{\pi}\frac{dp}{2\pi}\;\frac{w^{-\frac{1}{24}}}{f(w)}\sum_{n=-\infty}^{\infty} w^{(n+\beta(p)/4\pi)^2}
\end{equation}
The result (\ref{eq:R_infty}) provides the interpolation between the tight binding limit and the free string limit. 
Note that, for the critical boundary perturbation $g=\frac{1}{2}$, the rotation angle (\ref{eq:angle}) is $\beta = \pi$, independent of the momentum $p$. This verifies that the TBA is exact 
in this case, and the spectrum remains discrete for the noncompact $R=\infty$ theory.
For $g<\frac{1}{2}$ each of these states, which are infinitely degenerate in the $R=\infty$ theory, spreads into a band of width $1-2g$ around each integer. 
In the free string limit $g\rightarrow 0$, the gaps disappear and the bands merge to form the continuous dispersion relation 
of the free string.
In the field theory limit, the partition function reduces to that of the lowest band, $n=0$. For low momentum $p<<\pi$ we may expand the exponent in (\ref{eq:R_infty}),
\begin{equation}
\label{eq:lowestband}
\left(\frac{\beta(p)}{4\pi}\right)^2 - \frac{1}{24} = \left(\frac{g^2}{4}-\frac{1}{24}\right) + \left(\frac{g}{4\pi}\cot \pi g\right) \; p^2 + O(p^4)
\end{equation}
Thus in the continuum limit, we recover a relativistic dispersion relation with a mass
\begin{equation}
\label{eq:mass}
M^2 = \frac{1}{4\alpha'}\left(\frac{g^2}{4}-\frac{1}{24}\right)
\end{equation}
There is a range of subcritical values of the boundary perturbation,  $\frac{1}{\sqrt{6}}<g<\frac{1}{2}$, for which the tachyon has disappeared and the lowest lying mass squared value is positive. 
[When the confining force of the gauge field is included, the physical states are not charged $Z$ particles, but neutral $Z$ pairs, so $M$ is analogous to the constituent quark mass in QCD,
with the physical mass gap being $\approx 2M$. The stringy dynamics of the gauge field is discussed in the next Section.] 

The tachyon instability that we have found in the string-cutoff $CP^{N-1}$ model can be identified as the same instability found in the large $N$
approximation of the field theory with momentum space cutoff. A central result of the large $N$ analysis is the spontaneous generation of a mass term for the $Z$-particles. This results from
the fact that the equation obtained from variation of the action with respect to the Lagrange multiplier field $\lambda(x)$ allows a solution for nonzero constant $\lambda$, i.e
a finite $Z$ mass.  The equation which determines the mass is the one-loop tadpole condition that enforces $\langle Z^{\dag} Z\rangle = 1$, 
\begin{equation} 
\label{eq:gapeq}
1 = -i\gamma^2\int\frac{d^2p}{(2\pi)^2}\frac{1}{p^2+M^2}
\end{equation}
where $M^2 \equiv \gamma^2\lambda/N$. Introducing a momentum space cutoff $\Lambda$ in the loop integral, we get a positive mass squared
\begin{equation}
\label{eq:dimensional}
M^2 = \Lambda^2e^{-4\pi/\gamma^2}
\end{equation}
This result is a familiar example of dimensional transmutation, in which the physical scale is related to the cutoff scale by asymptotic freedom, as is the case in 4D QCD.
A similar result is obtained in the string-theoretic cutoff scheme, where $M^2$ is given by (\ref{eq:mass}),
Thus the tadpole condition (\ref{eq:gapeq}) determines the strength of the boundary perturbation describing the physical vacuum. 

The marginal boundary perturbation (\ref{eq:marginal}) is seen to be a generalization of the nonzero VEV acquired in the large N approximation by the Lagrange multiplier field $\lambda(x)$.
However, in the string theory case $\lambda$ is not a constant, but is rapidly oscillating in one spatial direction at the cutoff scale. In the continuum limit,
this rapidly oscillating field averages out to a net positive $Z$-particle mass shift. This is due to the fact that the $Z$-particle density also oscillates at the same rate, 
with the $Z$'s living preferentially in the 
vicinity of the D-branes. From Eq. (\ref{eq:lowestband}) we see that the effect of the boundary perturbation on the lowest lying $Z$ band is an overall positive 
mass shift, just like the effect of the constant $\lambda$ in the large $N$ calculation. The elegance of the particular marginal perturbation (\ref{eq:marginal}) is that the mass
shift appears through a shift of the Kac-Moody charge of the open string, thereby preserving world sheet conformal symmetry.

To summarize the main result of this section, the effect of the marginal boundary perturbation in the string-theoretic formulation of the $CP^{N-1}$ model is to induce a uniform shift 
in the Kac-Moody charges of the string states in the compactified $R=R_0$ theory, which is seen as a mass shift of the states in the $R=\infty$ theory. 
For the lowest lying band, the shift to a near-critical value of $g$ eliminates the tachyonic instability. This shift 
represents the transition from the perturbative vacuum to the physical vacuum. The critical boundary perturbation $g=\frac{1}{2}$ changes the boundary
condition on the string from Neumann to Dirichlet, and corresponds to a periodic background of codimension-one D-branes \cite{Gaiotto03,Lambert07}.
This picture of the physical vacuum nicely coincides with the codimension one, laminated topological charge structure observed in Monte Carlo simulations.

\section{Open-string closed-string duality}

It is interesting to view the tadpole condition (\ref{eq:gapeq}) in the context of open string/closed string (OS/CS) duality. This duality property is encoded
in the modular structure of the string loop amplitude and represents the fact that the evolution of the string worldsheet may be viewed in two equivalent ways. What has
been described here so far is the open string viewpoint, in which the $Z$ particle is replaced by a short piece of string which propagates in a closed loop with Neumann boundary
conditions on the string coordinates $X^{\mu}$ interacting with background $\lambda$ and $A_{\mu}$ fields. 
But we can also view the amplitude as describing the propagation of a closed string between two boundary states. The field theory limit from this viewpoint is the limit of vanishing
propagation time between the two closed string boundary states and the vacuum amplitude reduces to the overlap between the boundary states. In this way, one may
interpret the two sides of the large $N$ gap equation (\ref{eq:gapeq}) as the field theoretic limit of the equation which identifies the open and closed string representations
of the vacuum amplitude. Defining
\begin{equation}
Z(w,\beta) = \frac{w^{-\frac{1}{24}}}{f(w)}\sum_{n=-\infty}^{\infty}w^{(n+\beta/4\pi)^2}
\end{equation}
we can write the string generalization of the large $N$ gap equation (\ref{eq:gapeq}) as
\begin{equation}
\label{eq:string_gapeq}
1 = -i\gamma^2\int_0^{\infty}d\tau \int\frac{d^2p}{(2\pi)^2} Z(e^{-\tau/4\alpha'},\beta(p_1)) e^{-\tau p_2^2}
\end{equation}
Actually, from the OS/CS viewpoint the ultraviolet divergent ($\tau\rightarrow 0$)  part of the integrand on the right hand side of (\ref{eq:string_gapeq}) should be moved to the left hand side, 
since it is properly interpreted as coming from infrared behavior of the closed string channel, using the OS/CS duality transform (\ref{eq:OSCS}).

The closed string view provides additional insight into the nature of the excitations in the tachyonic crystal vacuum. While the open string represents the $Z$-particle in the 
field theory limit, the closed string states which arise by a dual transformation of the open string loop are associated with gauge excitations and Wilson loop operators in the
field theory. The boundary states in the closed string channel correspond to different configurations of D-branes in the spacetime background. Low lying excitations in this channel
describe collective transverse modes of the codimension-one D-branes in the tachyonic crystal and are thus associated with topological charge fluctuations in the field theory.
For finite cutoff $\alpha'^{-1}$, the ultraviolet limit of the loop integrand in (\ref{eq:string_gapeq}) is controlled by the limit $q\rightarrow 0$ ($w\rightarrow 1$), i.e. by the 
low energy or long range limit of the closed string channel. This is an example of a general feature of string theory, that ultraviolet divergences can be reinterpreted as 
infrared divergences in the dual channel. In the next section, we will investigate the effective action of the gauge field $A_{\mu}$ coupled to the open string as in Eq. (\ref{eq:boundary}).

\section{The effective gauge action: tachyonic crystals and quark confinement}

\begin{figure}
\vspace*{1.0cm}
\includegraphics{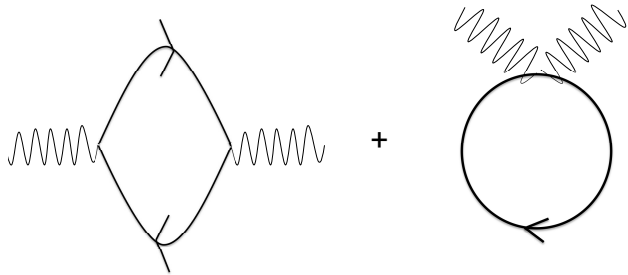}
\vspace{4.0cm}
\caption{One loop diagrams contributing to the large $N$ effective gauge potential in $CP^{N-1}$.}
\label{fig:oneloop}
\end{figure}

A central feature of the large N solution in $CP^{N-1}$ is the appearance of a dynamically generated linear confining Coulomb potential between charged $Z$-particles. This
arises from the fact that, although the bare Lagrangian does not include an $F^2$ term, such a term appears in the effective gauge potential at the one-loop level. This is seen by computing 
the one-loop photon self energy given by the diagrams in Fig. 3 (with $Z$-propagators of mass $M$).
In the limit $q^2\rightarrow 0$, this gives a self-energy tensor of the form
\begin{equation}
\Pi_{\mu\nu} = \frac{1}{4\pi^2\chi_t}(q^2g_{\mu\nu}-q_{\mu}q_{\nu})
\end{equation}
with $\chi_t = 3M^2/\pi N$.
This corresponds to a one-loop contribution to the effective gauge potential, 
\begin{equation}
{\cal L}_{eff} = -\frac{N}{48\pi^2}F_{\mu\nu}^2
\end{equation}
This promotes the auxiliary gauge field $A_{\mu}$ to a full dynamical Maxwell field, and in particular, leads to a linear confining potential between charged $Z$'s. 

In this Section we discuss the effective gauge potential in the string theoretic cutoff theory. Thus we consider an open string coupled to a background abelian
gauge field by the Wilson line interaction, Eq. (\ref{eq:boundary}). This problem has been studied in the context of boundary conformal field theory in
Ref. \cite{Callan87}. To present a self-contained discussion, we briefly summarize the results needed to obtain the effective gauge potential for the $CP^{N-1}$ model.
The analysis of Ref. \cite{Callan87} relies on an essential idea in the dynamics of strings in background fields. That is that the space-time equations of motion for the background
field are given by the requirement that the Wilson line boundary term in the BCFT preserves the conformal invariance of the world sheet theory. To determine the spacetime equations of motion for
the gauge field, we compute the CFT beta function associated with the boundary term and set it equal to zero. The appropriate beta function is defined by introducing a renormalization
counterterm for the Wilson line boundary interaction of the form
\begin{equation}
\Delta S = \frac{i}{2\pi}\int_{\sigma=0}d\tau \Gamma_{\mu}\partial_{\tau}X^{\mu}
\end{equation}
In the CFT calculation of the counterterm, an ultraviolet cutoff $\Lambda$ must be introduced.
The beta function which determines the conformal symmetry of the boundary term is then
\begin{equation}
\beta_{\mu} = \Lambda\frac{\partial}{\partial\Lambda}\Gamma_{\mu}
\end{equation}
The requirement that this beta function should vanish leads to the spacetime gauge field EOM,
\begin{equation}
\label{eq:EOM}
(\nabla^{\nu}\hat{F}_{\mu}^{\;\;\lambda})(1-\hat{F}^2)_{\lambda\nu}^{-1} = 0
\end{equation}
where $\hat{F}\equiv 2\pi\alpha' F$ and $F_{\mu\nu}=\partial_{\mu}A_{\nu}-\partial_{\nu}A_{\mu}$. 
To leading order in $\alpha'$, this is just Maxwell's equation. In fact, the exact equation of motion (\ref{eq:EOM}) can be 
derived from an effective gauge action of the Dirac-Born-Infeld (DBI) form,
\begin{equation}
\label{eq:DBI}
{\cal L}_{eff} \propto \sqrt{{\rm det}(1+\hat{F})}
\end{equation}
In the $\alpha'\rightarrow 0$ limit, this coincides with the large $N$ result ${\cal L}_{eff}\propto Tr F^2$. But the fact that the full action in the string-theoretic formulation
is of the DBI form has very interesting implications for the dynamics of electric flux-string formation \cite{Callan98}. Since the Coulomb force in two dimensions is linearly confining
even in perturbation theory, one might not expect the confinement mechanism exhibited by the $CP^{N-1}$ model to be particularly illuminating for 4-dimensional QCD. But 
the connection between 2- and 4-dimensional confinement appears to be much closer if we
view confinement in the $CP^{N-1}$ model in the context of brane dynamics in a tachyonic crystal vacuum. Electric flux tubes are associated with transverse motion of the branes (since they
are fundamental strings). Thus the spatial dimensionality of the electric flux distribution between charges is
determined by the codimensionality of the branes, suggesting a straightforward generalization of this mechanism to 4-dimensional QCD. As discussed in Ref. \cite{Callan98} linear strings of electric
flux arise from the DBI action by considering a point-charge source for a gauge field living on a D-brane. In order to construct a BPS configuration, the source must be accompanied by a transverse
fluctuation of the brane localized in the vicinity of the point source. This spike-like excitation can be interpreted as a fundamental string attached to the brane, or equivalently a tube
of electric flux attached to the source. The fact that the transverse excitation is string-like and highly localized in the brane coordinates depends crucially on the specific nonlinearities 
of the DBI gauge action \cite{Callan98}. Similarly, one may construct a BPS configuration with a transverse string passing through the brane. Again the DBI dynamics will constrict the
region through which the flux passes to a narrow tube. This analysis suggests that a tachyonic crystal of codimension-one D-branes will support string-like excitations which are stretched along the direction 
transverse to the local orientation of the branes. 

\section{Discussion}

In this paper I have explored a scenario for the breakdown of the perturbative vacuum in asymptotically free gauge theory and the appearance of the physical 
vacuum as a laminated array of codimension-one membranes. The evidence that such a lamination of the vacuum occurs in 4-dimensional $SU(3)$ Yang-Mills theory first appeared
in Monte Carlo studies of topological charge \cite{Horvath03}. In holographic QCD, the relevance of D6-branes for topological charge dynamics was first pointed out by Witten \cite{Witten98}.
This work showed that, in large $N$ HQCD, the basic objects that should appear in the vacuum as the carriers of topological charge should be codimension-one membranes rather
than localized, quantized instantons. The observation of codimension-one topological charge membranes in Monte Carlo configurations provided
a direct confirmation of Witten's arguments. In addition, the Monte Carlo results showed that, not only does topological charge appear in the form of thin, extended, coherent membranes, but that
the vacuum was essentially filled with a topological sandwich of closely spaced, alternating sign membranes. The central argument of this paper is that the observed alternating-sign
sandwich array has a precise holographic interpretation as a tachyonic crystal, the expected decay product of a homogeneous, space-filling D7-brane. This presents an appealing scenario
for the breakdown of the perturbative vacuum and the transition to the physical one by a lamination process that can be described in terms of an exactly marginal boundary perturbation
in the conformal field theory of the coordinate along which the lamination occurs. In this respect, the local description of the vacuum lamination phenomenon should be very similar for
2-dimensional $CP^{N-1}$ and 4-dimensional QCD, since both dynamics are described by the BCFT for a single spatial coordinate transverse to the array of codimension-one D-branes. 

To provide further support for the tachyon crystal model of the gauge theory vacuum, we presented a detailed discussion of this scenario in the $CP^{N-1}$ model, which provides a two-dimensional
model with a topological charge structure very similar to QCD. For our discussion of the $CP^{N-1}$ model, the full superstructure of a gauge/gravity dual was
not necessary to arrive at the 
BCFT which describes the tachyonic decay of the vacuum. Instead we showed that the $CP^{N-1}$ model could be formulated as the field theoretic limit of an open bosonic string in 
background fields $\lambda$ and $A_{\mu}$. In a proper time formulation, the propagator for the charged field $Z$ is given by the $\alpha'\rightarrow 0$ limit of the open string propagator.
The string length parameter $\sqrt{\alpha'}$ plays a role analogous to lattice spacing in a lattice-regulated gauge theory. This string-theoretic cutoff naturally incorporates one of the most striking features
of the observed topological charge sheets on the lattice, the fact that the thickness of each sheet and the separation between sheets is determined by the cutoff scale and approaches zero in the continuum limit.

An important issue which has not been adequately addressed in this paper is the question of scaling and universality in a theory whose properties depend so crucially on dynamics that occurs near the cutoff scale.
In particular, the proper scaling behavior of the topological susceptibility is not a foregone conclusion. As an example of what can go wrong, the simplest lattice formulation of the
$CP^1$ and $CP^2$ models do not exhibit proper scaling for $\chi_t$ due to small instantons\cite{Berg,Lian07}, and it is not
clear whether the topological susceptibility in these models can be defined as the continuum limit of a lattice calculation. 
Fortunately, for both $CP^{N-1}$ with $N>4$ \cite{Ahmad05} and for 4-dimensional $SU(3)$ gauge
theory \cite{Bardeen04}, small instantons do not contribute, and $\chi_t$ is found to scale properly. 
So at least the numerical evidence indicates that the topological sandwich structure leads to a finite topological susceptibility in the continuum limit and
does not cause the same scaling difficulties as small instantons. It is interesting to note that as the scaling limit is approached 
the scaling of $\chi_t$ takes place by virtue of an increasingly precise cancellation in the integrated correlator between the positive contribution at the origin and the negative tail
at $|x|\geq 2$ lattice spacings, with the positive and negative contributions separately diverging in the continuum limit \cite{Horvath_corr,Ahmad05}. Since the positive contribution comes from each
individual brane, while the negative contribution comes from the juxtaposition of oppositely charged branes, this cancellation is strong evidence that the
positions of the layered membranes are highly correlated and arranged in a regular periodic array, as would be expected in the tachyon crystal scenario.
We can conclude from this that the presence of thin membranes of topological charge in the vacuum does indeed pose a threat to proper scaling, just as small instantons do,
in the sense that a random distribution of such membranes would lead to a divergent $\chi_t$..
It is only through the regular ``antiferromagnetic'' ordering of the membranes that proper scaling is achieved. 

The issue of universality (i.e. whether different cutoffs lead to the same low energy physics) may be further illuminated in the $CP^{N-1}$ model by detailed comparison
between the string theory cutoff, the lattice cutoff, and the large N approximation with momentum space cutoff. In fact an apparent discrepency has been pointed out between the lattice
formulation and the continuum large $N$ approximation associated with the phenomenon of ``superconfinement'' \cite{Rabinovici81,Samuel83}. This is the observation that, to all orders
in the lattice strong coupling expansion, the net $U(1)$ current (\ref{eq:U1current}) flowing along any link of the lattice must be zero, since the integral over the gauge link vanishes unless there
are an equal number of $U$'s and $U^{\dag}$'s. This would seem to rule out bulk electric field excitations.
What appears in the large N approximation as a constant electric field, e.g. between spatially separated test charges, when looked
at microscopically on the lattice is a chain of polarized $Z^+Z^-$ pairs.  The large $N$ approximation thus gives a somewhat smeared out view of the short-distance lattice dynamics.
On the other hand, lattice Monte Carlo calculations of $\chi_t$ converge nicely with increasing N to be in accurate agreement with the large N result at about $N\approx 9$ \cite{Lian07}. So the
absence of superconfinement in the large $N$ approximation does not seem to invalidate its low energy predictions.
As discussed in \cite{Samuel83} the smearing of the superconfinement constraint and the appearance of bulk electric fields in the large $N$ analysis can be traced to the introduction of the Lagrange multiplier
field $\lambda$, which temporarily relaxes the constraint $Z^{\dag}Z=1$. This constraint is reimposed upon integrating over the $\lambda$ field, but for fixed $\lambda$,
superconfinement does not hold. (Note that typically the Lagrange multiplier field is not introduced in lattice calculations and the constraint $Z^{\dag}Z=1$ is maintained site by site 
in the update procedure.) The large $N$ approximation, in which $\lambda$ is assumed to be a constant, is not sufficient to restore the superconfinement property.
It is very interesting that the string theoretic cutoff introduced in this paper leads to a $\lambda(x)$ field which is rapidly oscillating at the cutoff scale but has the
same low energy effect as a constant $\lambda$, i.e. a constant mass shift. The tachyonic crystal structure for the gauge field which emerges from the string theory 
seems to be to some extent replicating the short-distance structure seen on the lattice.

This work was supported in part by the Department of Energy under grant DE-FG02-97ER41027.

\end {document}